\documentclass[10pt, notitlepage, twocolumn, superscriptaddress, nofootinbib, floatfix]{revtex4-1}

\usepackage{amsmath}
\usepackage{mathrsfs}
\usepackage{multirow}
\usepackage{amssymb}
\usepackage{mathtools}
\usepackage{hyperref}
\usepackage[usenames,dvipsnames]{xcolor}
\hypersetup{colorlinks=true, citecolor=Purple, linkcolor=Purple,
urlcolor=Purple}

\usepackage{graphicx}
\usepackage{orcidlink}

\newcommand{\GB}{\mathscr{G}}

\usepackage{tikz}
\usepackage[most]{tcolorbox}
\usepackage{empheq}
\usetikzlibrary{shadows}
\tcbset{
  highlight math style={
    enhanced,
    colframe=blue!40!black,
    colback=blue!10,
    arc=0pt,
    boxrule=1pt,
    drop shadow
  }
}

\begin{document}

\allowdisplaybreaks

\title{Gravitational quasinormal modes of black holes in quadratic gravity}

\author{Georgios~Antoniou}
\email{georgios.antoniou@roma1.infn.it}
\affiliation{Dipartimento di Fisica, ``Sapienza'' Universit\'a di Roma, P.A. Moro 5, 00185, Roma, Italy}
\affiliation{Sezione INFN Roma1, P.A. Moro 5, 00185, Roma, Italy}
\affiliation{Dipartimento di Fisica,  Universit\`a di Pisa, 56127 Pisa, Italy}

\author{Leonardo~Gualtieri}
\email{leonardo.gualtieri@unipi.it}
\affiliation{Dipartimento di Fisica,  Universit\`a di Pisa, 56127 Pisa, Italy}
\affiliation{INFN, Sezione di Pisa, Largo B. Pontecorvo 3, 56127 Pisa, Italy}

\author{Paolo~Pani}
\email{paolo.pani@uniroma1.it}
\affiliation{Dipartimento di Fisica, ``Sapienza'' Universit\'a di Roma, P.A. Moro 5, 00185, Roma, Italy}
\affiliation{Sezione INFN Roma1, P.A. Moro 5, 00185, Roma, Italy}

\begin{abstract}    
    We study the gravitational perturbations of black holes in quadratic gravity, in which the Einstein-Hilbert term is supplemented by  quadratic terms in the curvature tensor. 
    In this class of theories, the Schwarzschild solution can coexist with modified black hole solutions, and both families are radially stable in a wide region of the parameter space.
    Here we study non-radial perturbations of both families of static, spherically symmetric black holes, computing the quasi-normal modes with axial parity and finding strong numerical evidence for the stability of these solutions under axial perturbations. The perturbation equations describe the propagation of a massless and a massive spin-two fields.
    We show that the Schwarzschild solution admits the same quasi-normal modes as in general relativity, together with new classes of modes corresponding to the massive spin-two degrees of freedom. The spectrum of the  modified black hole solution has the same structure, but all modes are different from those of general relativity.   
    We argue that  both classes of modes can be excited in physical processes, suggesting that a characteristic signature of this theory  is the presence of massive spin-two modes in the gravitational ringdown, even when the stationary solution is the same as in general relativity.
\end{abstract}

\maketitle

\section{Introduction}
Over the past few years, gravitational-wave~(GW) astronomy has rapidly developed, proving its potential through the many GW events observed so far~\cite{LIGOScientific:2016aoc,LIGOScientific:2018mvr,LIGOScientific:2020ibl,LIGOScientific:2021usb,KAGRA:2021vkt}. 
GW signals from binary mergers have been extensively used to test General Relativity~(GR) in the strong-gravity regime~\cite{LIGOScientific:2021sio} and provide some of the most stringent bounds on alternative theories of gravity~\cite{Berti:2015itd,Yunes:2024lzm}. The accuracy of these tests of gravity will improve as more events are detected in the following years and with the advent of next-generation interferometers on ground~\cite{Maggiore:2019uih,Kalogera:2021bya,Branchesi:2023mws} and in space~\cite{LISA:2022kgy,Barausse:2020rsu,Maggiore:2019uih,Kalogera:2021bya,Colpi:2024xhw}.

Arguably, the most natural way to modify GR is by introducing higher derivative operators to the Einstein-Hilbert action~\cite{Berti:2015itd}. Such terms may be motivated either from an effective field~(EFT) theory point of view or from more fundamental arguments. For example, in so-called quadratic gravity the gravitational action is modified by including terms that are quadratic in the curvature tensors. Such terms make the theory formally renormablizable~\cite{Stelle:1976gc} but introduce ghosts, associated with extra degrees of freedom of massive spin-0 and spin-2 modes propagating in this theory.

These ghosts may challenge the viability of quadratic gravity as a fundamental theory, even though this is presently matter of debate (see e.g.\,\cite{Donoghue:2021cza}). 
On the other hand, it can be shown (see e.g.\,\cite{Burgess:2003jk,Endlich:2017tqa}) that by considering quadratic gravity as an EFT, the deviations from GR can be transformed by a field redefinition to higher-order terms.

Despite the presence of ghosts and the fourth-order field equations stemming from its action, quadratic gravity was found to lead to a well-posed initial-value formulation~\cite{Noakes:1983xd}.
This has motivated recent phenomenological and numerical studies on this theory beyond the EFT approach, at least at the classical level~\cite{Lu:2015cqa,Lu:2015psa,Kokkotas:2017zwt,Konoplya:2022iyn,East:2023nsk
,Held:2023aap}.
Furthermore, certain subclasses of quadratic gravity have gathered particular interest in the context of inflationary cosmology (e.g., Starobinski's model of inflation~\cite{Starobinsky:1980te}) since quadratic curvature corrections might be relevant in the early universe.

In this paper we will take the same perspective of recent studies~\cite{Lu:2015cqa,Lu:2015psa,Kokkotas:2017zwt,Konoplya:2022iyn,East:2023nsk,Held:2021pht,Held:2023aap}, and consider quadratic gravity as a full theory. When needed, we will highlight the differences and the common regimes of validity with the EFT. As we are going to discuss, some of our results are also relevant for an analysis in an EFT framework, possibly at the same order as quartic terms in the curvature tensor such as those studied in\,\cite{Endlich:2017tqa,Cardoso:2018ptl}.

A natural arena to explore the implications of quadratic curvature terms are black holes (BHs)~\cite{Barack:2018yly}. As long as we consider perturbations of static, asymptotically flat spacetimes, it can be shown\,\cite{Nelson:2010ig,Lu:2015cqa} that it is not restrictive to consider a subclass of quadratic gravity theories, the so-called Einstein-Weyl~(EW) gravity, in which the Einstein-Hilbert term is supplemented by the square of the Weyl tensor. This term has a dimensionful coupling constant $\alpha$, which is associated with the mass $\mu$ of the massive spin-2 degree of freedom propagating in this theory, namely $\alpha\equiv 1/(2\mu^2)$ (hereafter we use $G=c=1$ units). The GR limit wherein an EFT approach is valid corresponds to $\alpha\to0$ or, equivalently, $\mu\to\infty$, where the extra modes becomes very heavy.

Within GR, BH uniqueness theorems (see e.g. \cite{Chrusciel:2012jk,Cardoso:2016ryw} and references therein)  mandate that the Schwarzschild metric is the only  spherically symmetric, asymptotically flat, vacuum solution. EW gravity, instead, predicts the presence of a non-GR branch coexisting with the Schwarzschild solution. The Weyl term modifies the equations of motion, giving rise to spherically symmetric and asymptotically flat BHs beyond GR.
For simplicity, we will refer to the latter family of solutions as ``hairy'' BHs, although all the degrees of freedom in this theory are purely gravitational.

The horizon radius $r_h$ of the hairy BH is bounded both from above and from below.
BH solutions with $r_h$ larger than the upper bound have negative mass and are thus unphysical.
Indeed, at variance with the Schwarzschild solution, the mass of the hairy BH is a \emph{decreasing} function of the horizon radius. 
BH solutions with $r_h$ smaller than the lower bound are unstable against radial perturbations~\cite{Held:2022abx}.
In terms of the dimensionless parameter $p\equiv r_h/\sqrt{2\alpha}$, radially stable hairy BHs with a positive mass exist for $0.876 \lesssim p \lesssim 1.143$. 
The upper bound on $p$ in the domain of existence suggests that hairy BHs are non-perturbative solutions which do not exist within the EFT.

In the case of stationary, rotating BHs, an approximate computation shows the same structure of the solution space: the Kerr solution is stable above a critical value of mass, while a hairy BH solution exists only below this value\,\cite{Sajadi:2023smm}.

In this paper we consider non-radial perturbations of both branches of spherically symmetric BH solutions in EW gravity,  studying its
quasi-normal modes (QNMs) of oscillation, which describe the damped sinusoids corresponding to the free oscillations of a perturbed BH.

The aim of this study if twofold. Firstly, we want to study the non-radial stability of BHs in EW gravity and thus, more generally, in quadratic gravity. Curiously, for $0.876 \lesssim p \lesssim 1.143$ {\it both} the Schwarzschild BH and the hairy BH solutions are radially stable, but this does not guarantee that they are stable for non-radial perturbations as well. Examining the stability of BHs within such extended theories is crucial for assessing their physical viability.

Secondly, the QNM spectrum is a promising probe to observationally test if GR deviations of the class of those studied in this paper are actually present. 
QNMs play a crucial role in the post-merger ringdown phase at the end of a binary BH coalescence and  encapsulate important information about the underlying gravitational theory (see~\cite{Berti:2018vdi} for a review). Today, QNMs have been measured for a number of binary BH coalescences\,\cite{Ghosh:2021mrv,LIGOScientific:2021sio}. 
In GR, the QNMs have been extensively studied and have been shown to be determined by the mass, charge, and spin of the BH, in accordance with no-hair theorems. 
However, in theories beyond GR the ringdown spectrum exhibits distinct features that can be used to distinguish them  from GR (see~\cite{Pani:2009wy,Cardoso:2009pk,Molina:2010fb,Pani:2013ija,Pani:2013wsa,Blazquez-Salcedo:2016enn,Blazquez-Salcedo:2020rhf,Blazquez-Salcedo:2020caw,Pierini:2021jxd,Wagle:2021tam,Cano:2021myl,Pierini:2022eim,Cano:2023jbk,Wagle:2023fwl,Antoniou:2024gdf,Antoniou:2024hlf,Cano:2024ezp,Chung:2024ira,Chung:2024vaf} for recent progress in computing QNMs in theories beyond GR). 
In particular, any modified gravity theory generically introduces two modifications to the ringdown: i)~a deformed QNM spectrum, whose deviations from the GR case are proportional to (powers of) the coupling constant; ii)~novel extra modes in the gravitational waveform~\cite{DAddario:2023erc,Crescimbeni:2024sam}, whose amplitude is proportional to (powers of) the coupling.

The complexity of the equations derived in Einstein-Weyl gravity poses significant difficulty in performing the ringdown analysis.
It has been shown that also the Schwarzschild branch presents a monopolar instability for values $p\lesssim 0.876$, i.e. for small mass solutions~\cite{Babichev:2013una,Brito:2013wya} (see also~\cite{Myung:2023ygn,Myung:2013doa}). This result is equivalent to the Gregory-Laflamme instability~\cite{Gregory:1993vy} for high-dimensional black strings~\cite{Babichev:2013una}. More recently, the monopolar stability of the non-GR branch was examined and a similar instability was discovered in the same range $p\lesssim 0.876$~\cite{Held:2022abx,Held:2023aap}. 

BHs for  $p>0.876$, instead, are stable for radial perturbations. Time-domain evolutions of BHs in the Schwarzschild branch\,\cite{Held:2021pht}, give strong indications for their non-radial stability (see also\,\cite{Held:2023aap} for the evolution of rotating BHs and of BH binaries).
Moreover, scalar and electromagnetic perturbations of BHs in the non-GR branch have been studied using approximate techniques, finding stable modes\,\cite{Cai:2015fia,Zinhailo:2018ska,Zinhailo:2019rwd,Konoplya:2022iyn}. In this paper we compute for the first time gravitational QNMs, in the axial sector, for both branches of spherically symmetric BH solutions in EW gravity, finding as expected a deformation of the GR spectrum and new classes of modes. 

Moreover, we study whether the new modes can be excited in the ringdown stage of a binary BH coalescence. Then, using the results of our study, we discuss the phenomenological perspectives of EW gravity, i.e. whether we expect the non-GR features of BHs in this theory, if present, could be observed by present or near-future GW detectors. 

The paper's structure is the following: in Sec.\,\ref{sec:framework} we present the theoretical framework of the model. In Sec.\,\ref{sec:bg} we review the background solutions in this theory, using both numerical and  semi-analytical treatments. In Sec.\,\ref{sec:perturbations} we analyze the axial perturbations in EW gravity, first on a Schwarzschild background recovering the results presented in~\cite{Brito:2013wya}. We show our analysis for the axial QNMs of the non-GR branch of solutions in Sec.\,\ref{subsec:R_flat}.
In Sec.\,\ref{sec:pheno} we discuss the phenomenological perspectives of EW gravity. Finally, we draw our conclusions in  Sec.\,\ref{sec:conclusions}.

\section{Framework}
\label{sec:framework}
In four-dimensional gravity, the most general theory involving only the Einstein-Hilbert term and quadratic curvature invariants is given by~\cite{Stelle:1976gc,Stelle:1977ry,Whitt:1985ki}
\begin{equation}
    \mathcal{I}=\int {\rm d}^4 x \sqrt{-g} \left( R-\alpha C_{\mu \nu \rho \sigma} C^{\mu \nu \rho \sigma}+\beta R^2 \right)\,,
\label{eq:action}
\end{equation}
where $\sqrt{-g}$ is the determinant of the metric tensor, $R$ is the Ricci tensor and $C_{\mu \nu \rho \sigma} C^{\mu \nu \rho \sigma}$ is the square of the Weyl tensor.
The latter can be expressed in terms of the Ricci and Gauss-Bonnet (GB) invariants in the following way
\begin{equation}
    C_{\mu \nu \rho \sigma} C^{\mu \nu \rho \sigma}=2R_{\mu \nu} R^{\mu \nu}-\frac{2}{3}R^2+{\GB}\,,
\label{eq:Weyl_scalar}
\end{equation}
where the GB invariant is defined as $\GB=R^2-4R_{\mu \nu}R^{\mu \nu}
+R_{\mu \nu \rho \sigma}R^{\mu \nu \rho \sigma}$.
The dimensionful parameters $\alpha$ and $\beta$ are associated with additional massive degrees of freedom. Specifically, in addition to the massless spin-2 graviton, this theory predicts a massive spin-0 mode with mass $m^2=1/(6\beta)$, and a massive spin-2 mode with mass $\mu^2=1/(2\alpha)$.
Remarkably, the sign of the kinetic term of the massive spin-2 mode is the opposite than those of the other kinetic terms, i.e. it is a ghost-like degree of freedom. This suggests that quadratic gravity is non-unitary as a full quantum theory (however, see~\cite{Donoghue:2021cza} for different interpretations); 
this problem is avoided by considering quadratic gravity as a low-energy EFT, as long as the energy scale is lower than a cut-off scale which is of the order of $\alpha^{-1/2}\sim\mu$. However, in this framework the actual deviations from GR are expected to occur at higher order in the EFT expansion, since the quadratic corrections can be transformed by a field redefinition to higher-order terms\,\cite{Burgess:2003jk,Endlich:2017tqa}.

Note that the existence of the hairy BH solution, occurring for $p=r_h\mu\sim 1$ and then at energy scales close to $\mu$, is questionable in an EFT framework. 
For completeness, in this work we will compute the QNMs of both the Schwarzschild BH solution (which clearly exists also in the EFT) and of the hairy BH solution (which instead exists only in the full theory).
We also remark that the existence of an Ostrogradsky ghost does not prevent the theory from having a well-posed initial value formulation at a classical level~\cite{Noakes:1983xd,Held:2021pht}. 

As discussed in~\cite{Nelson:2010ig,Lu:2015cqa}, static, asymptotically flat solutions of quadratic gravity have vanishing Ricci scalar, and thus the term $\beta R^2$ does not affect the linear perturbations of such solutions. Therefore, without loss of generality we may neglect the $\beta R^2$ term in our analysis, setting $\beta=0$, and focussing on the pure EW gravity.

Variation of the action with respect to the metric tensor yields the field equations
\begin{align}
    &
    G_{\mu\nu}+\alpha\big[
    -4R_\mu{}^\rho R_{\nu\rho}+g_{\mu\nu}(R_{\rho\sigma})^2+(2/3)\big(R_{\mu\nu} \nonumber\\
    &
    +g_{\mu\nu}\nabla^2+ \nabla_\mu \nabla_\nu \big)R+4\nabla_\rho \nabla_{(\mu}R_{\nu)}{}^\rho-2\nabla^2 R_{\mu\nu} \label{eq:field_high_order}\\
    &
    -g_{\mu\nu}R^2/6-2g_{\mu\nu}\nabla_\sigma\nabla_\rho R^{\rho\sigma}+2R_\mu{}^{\rho\sigma\lambda}R_{\nu\rho\sigma\lambda}\nonumber\\
    &
    -g_{\mu\nu}(R_{\rho\sigma\lambda})^2/2+4\nabla_{(\rho} \nabla_{\sigma)}R_\mu{}^\rho{}_\nu{}^\sigma\big]=0 \, , \nonumber
\end{align}
where beyond-GR contributions are proportional to a factor $\alpha$.
Despite not being obvious at first glance, the field equations for spherically symmetric configurations yield second-order equations. We will go over that in the following subsections.

\subsection{Auxiliary field}
Here we introduce an additional massive auxiliary field. This allows us to write the Lagrangian in a form leading to second-order field equations.

To begin with, we take advantage of the fact that the Weyl tensor can be expressed in terms of the GB invariant, which does not contribute to the action in 4 dimensions. Then, setting $\beta=0$ as discussed above, Eq.\,\eqref{eq:action} reads:
\begin{equation}
    \mathcal{I}=\int {\rm d}^4 x \sqrt{-g} \left[ R-\frac{1}{2\mu^2}\left(2R_{\mu \nu} R^{\mu \nu}-\frac{2}{3}R^2\right) \right]\,.
\label{eq:action_2}
\end{equation}
Following~\cite{Hindawi:1995an,Brito:2013wya,Hinterbichler:2015soa,Held:2022abx,Held:2023aap}, we introduce an auxiliary tensor field $f_{\mu\nu}$, rewriting the action as follows:
\begin{equation}
    \mathcal{I}_0=
        \int {\rm d}^4 x \sqrt{-g} \left[R + 2 f_{\mu\nu}G^{\mu\nu}
        +{\mu^2}\left(f_{\mu\nu}f^{\mu\nu}-f^2\right)\right]
    \label{eq:Lagrangian}
\end{equation}
The equations of motion for $g_{\mu\nu}$ can be found by taking variations of the action with respect to $f_{\mu\nu}$
\begin{equation}
     \mathcal{E}^{(g)}_{\mu\nu}\equiv G_{\mu\nu}+ \mu^2 \left(f_{\mu\nu}-f g_{\mu\nu}\right)=0\, ,
\label{eq:field_g}
\end{equation}
from which we find that
\begin{equation}
    R_{\mu\nu} = -\mu^2 \left(f_{\mu\nu}+\frac{1}{2} f g_{\mu\nu}\right)\;, \; R=-3\,\mu^2f\, ,
\label{eq:Ricci_f}
\end{equation}
or equivalently
\begin{equation}
    f_{\mu\nu} = -\frac{1}{\mu^2} \left(R_{\mu\nu}-\frac{1}{6}R \,g_{\mu\nu}\right)\, .
\label{eq:f_R}
\end{equation}
It is straightforward to confirm that substitution of the above into~\eqref{eq:Lagrangian} leads us back to~\eqref{eq:action_2}.
Furthermore, from the Bianchi identity we have $\nabla^\mu G_{\mu\nu}=0$ which, by making use of~\eqref{eq:field_g}, results in the constraint
\begin{equation}
    \mathcal{E}_\mu\equiv\nabla^\nu f_{\mu\nu}-\nabla_\mu f=0\, .
\label{eq:nabla_f}
\end{equation}
Therefore, the equations of motion for $f_{\mu\nu}$ read
\begin{equation}
\begin{split}
    &
    \mathcal{E}^{(f)}_{\mu\nu}\equiv
    G_{\mu\nu}+G_{(\mu}{}^\rho f_{\nu)\rho}-g_{\mu\nu}G^{\rho\sigma}f_{\rho\sigma}-fR_{\mu\nu}
    \\
    &
    +R f_{\mu\nu}+\Box f_{\mu\nu}+(\nabla_\mu\nabla_\nu-g_{\mu\nu})f
    -2\nabla_\rho\nabla_{(\mu}f_{\nu)}{}^\rho\\
    &
    +g_{\mu\nu}\nabla_\rho\nabla_\sigma f^{\rho\sigma}+\mu^2\big[\big(f^2-f_{\rho\sigma}f^{\rho\sigma}\big)/2\\
    &
    +2\big(f_\mu{}^\rho f_{\nu\rho}-f\,f_{\mu\nu}\big)\big]=0\,.
 \end{split}
\label{eq:field_f}
\end{equation}
\par\noindent
We remark that the GR limit corresponds to $\mu\to\infty$, i.e. $\alpha\to0$. In this limit, the auxiliary field $f_{\mu\nu}$ (Eq.\,\eqref{eq:f_R}) vanishes, and Eqs.\,\eqref{eq:field_f} reduce to Einstein's equations in vacuum. Conversely, in the $\mu\to0$ limit the theory becomes strongly coupled,  the action only has quadratic curvature terms, and EW gravity reduces to the pure Weyl gravity \cite{Weyl:1918ib}.

\section{Background solutions}
\label{sec:bg}
In this section we review the background solutions found in EW gravity, including those of the non-GR branch.
We are interested in static and spherically symmetric solutions, so we consider the following metric element (following the notation of~\cite{Lu:2015cqa,Kokkotas:2017zwt})
\begin{equation}
    d s^2=-A(r) d t^{2} +\frac{1}{B(r)} d r^{2}+ r^{2} \left( d \theta^{2}+\sin^{2}\theta\,d\varphi^{2} \right)\,.
\label{eq:metric}
\end{equation}
In order to derive the background equations we firstly take the trace of Eq.~\eqref{eq:field_f} which yields
\begin{align}
    2\big(\nabla_\mu\nabla_\nu f^{\mu\nu}-\Box f\big) - R = 0\,.
\end{align}
From Eq.,\eqref{eq:nabla_f}, using the symmetry of $f_{\mu\nu}$, we find $\nabla_\mu\nabla_\nu f^{\mu\nu}=\nabla_\mu\nabla_\nu f^{\nu\mu}=\Box f$. 
From the trace of~\eqref{eq:field_f} we find then $R=0$: the spherically symmetric solutions (in vacuum) of EW gravity are necessarily \emph{Ricci-scalar flat}. This also means that from~\eqref{eq:Ricci_f} the trace of the auxiliary field vanishes, \textit{i.e.} $f=0$.
From this property we derive the first equation for the metric functions, i.e.
\begin{equation}
\begin{split}
    2 r^2 A B A''=&-r A A' \left(r B'+4 B\right)+r^2 B A'^2\\
    &-4 A^2 \left(r B'+B-1\right)\, .
\label{eq:bg_1}
\end{split}
\end{equation}
The next step is to consider Eq.~\eqref{eq:field_f} which allows us to express the components of $f_{\mu\nu}$ in terms of the metric functions $A,\,B$. We simplify the expressions by substituting~\eqref{eq:bg_1}. Finally, we substitute $f_{\mu\nu}$ and~\eqref{eq:bg_1} into~\eqref{eq:field_g} and from the $(rr)$ component we find
\begin{equation}
\begin{split}
    &
    2 r^2 A^2 B (r A'-2 A) B''-B' \big[4 r A^3-r A B (r^2 A'^2\\
    &
    +2 r A A'+4 A^2)\big]-4 A^2 B \big[\mu ^2 r^3 A'+A (\mu ^2 r^2+2)\big]\\
    &
    -B^2 (r^3 A'^3-3 r^2 A A'^2-8 A^3)+3 r^2 A^3 B'^2\\
    &
    +4 \mu ^2 r^2 A^3=0\, .
\end{split}
\label{eq:bg_2}
\end{equation}
Alternatively we could have used the higher-order equations~\eqref{eq:field_high_order}.
First we eliminate $A^{(4)}$ by making use of the $(tt)$ component. We substitute $A^{(4)}$ in the radial derivative of the $(rr)$ equation and solve for $A^{(3)}$. We then substitute $A^{(3)}$ in the $(rr)$ equation and retrieve~\eqref{eq:bg_1} which allows us to solve for $A''$. We may then substitute $A'',\, A^{(3)},\, A^{(4)}$ in the radial derivative of the $(rr)$ equation to retrieve~\eqref{eq:bg_2}.

Given that the static and spherically symmetric solutions have necessarily $R=0=f$, we will divide the analysis of BH solutions in the following two cases: 
\begin{enumerate}
    \item \emph{Ricci-tensor flat solutions} for which ${R}_{\mu\nu}=0$ (which of course implies $R=0$ and also $f_{\mu\nu}=0$ through Eq.~\eqref{eq:f_R}). Since the equations coincide with GR equations in vacuum, in this case the BH background is simply the Schwarzschild metric.
    \item \emph{Ricci-scalar flat solutions} for which ${R}=0$ but $R_{\mu\nu}\neq0$. In this case $f_{\mu\nu}\neq0$, the background solution is not Schwarzschild, and we will refer to it as {\it hairy BH}. 
\end{enumerate}
Obviously in the first case the solution can be written in analytical form, while the in the second case it has to be found numerically, by solving the system of ordinary differential equations with suitable boundary conditions.

\subsection{Expansions at the horizon}
We expand the metric functions according to the following ansatz,
where we denote the location of the BH horizon as $r=r_h$:
\begin{align}
A(r) &= c\bigg[(r-r_h) + \sum_{n=2}a_n(r-r_h)^n\bigg]\, ,\label{eq:bg_horizon_1}\\
B(r) &= \sum_{n=1}b_n(r-r_h)^n\, .\label{eq:bg_horizon_2}
\end{align}
To find the coefficients in the expansions we substitute them in the background equations, which we then solve order by order. We show here the first few coefficients:
\begin{align}
    a_2 = \,&-\frac{2}{r_h}+\frac{\mu^2}{4 r_h b_1^2}+\frac{1}{r_h^2 b_1}-\frac{\mu^2}{4 b_1}\,,\\
    \begin{split}
        a_3 = \,& \frac{7\mu^4}{72 b_1^2}+\frac{\mu^2}{3  b_1^3 r_h^3}-\frac{20}{9 b_1 r_h^3}+\frac{\mu^4}{8 b_1^4 r_h^2}\\
        & -\frac{13\mu^2}{36 b_1^2 r_h^2}-\frac{2\mu^4}{9 b_1^3 r_h}+\frac{\mu^2}{36 b_1 r_h}+\frac{29}{9 r_h^2}\, ,
    \end{split}
\\[2mm]
    b_2 = \,& \frac{1}{r_h^2}+\frac{3\mu^2}{4}-\frac{3\mu^2}{4 r_h b_1}-\frac{2 b_1}{r_h}\, ,\\
    \begin{split}
        b_3 = \,& \frac{\mu^4}{72 b_1}+\frac{\mu^2}{3 b_1^2 r_h^3}-\frac{\mu^2}{8 b_1^3 r_h^2}-\frac{7\mu^2}{36 b_1 r_h^2}\\
        &\,+\frac{17 b_1}{9 r_h^2}+\frac{\mu^4}{9 b_1^2 r_h}-\frac{8}{9 r_h^3}-\frac{5\mu^2}{36 r_h}\,.
    \end{split}
\end{align}
The coefficients above are determined in terms of $b_1$ which serves as the parameter quantifying the deviation from GR. In the limit $\alpha\to\infty$ ($\mu\to 0$), the GR sector decouples and we formally recover the GR near-horizon expansion coefficients. 

\subsection{Expansions at infinity}
The solutions we are interested in are asymptotically flat.
Therefore, we solve Eqs.\,\eqref{eq:bg_1} and\,\eqref{eq:bg_2} as a series expansion in $1/r$, with $A(r),B(r)\to 1$ for $r\to\infty$, finding:
\begin{align}
    &A=1-\frac{2M}{r}+\tilde{c}\frac{e^{-\mu r}}{r}\cdots\,,\label{eq:bg_infinity_1}\\
    &B=1-\frac{2M}{r}+\tilde{c}\,(1+\mu r)\frac{e^{-\mu r}}{r}\cdots\,.\label{eq:bg_infinity_2}
\end{align}
where $c$ and $M$ are free parameters. 
\begin{figure}
    \centering
    \includegraphics{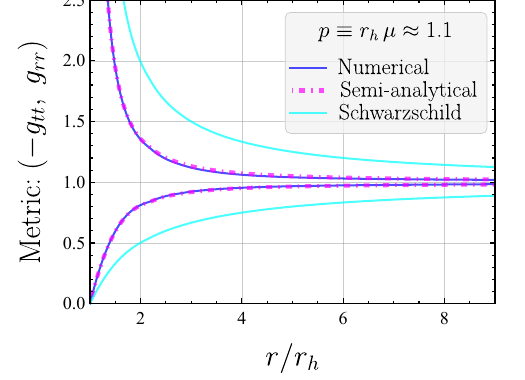}
    \caption{The metric components for a Schwarzschild BH and a hairy BH, with $p\equiv r_h\mu=1.1$, and with the same horizon radius.}
\label{fig:metric}
\end{figure}

\subsection{Semi-analytical background}
In\,\cite{Kokkotas:2017zwt}, using the parametrization introduced in\,\cite{Rezzolla:2014mua}, the authors derived a semi-analytical approximation of the static BH solutions in EW gravity, in terms of continued fractions. If we define $x\equiv (1-r_h/r)$, then
\begin{align}
    A(r) & \equiv x\,f(x)\,,\\
    \frac{A(r)}{B(r)} & \equiv h(x)^2\,,\label{eq:fh}
\end{align}
where the functions $f$ and $h$ are given in Appendix\,\ref{app:cfsolution} in terms of continued fractions.

In Fig.~\ref{fig:metric} we present, as an example, the $(tt)$ and $(rr)$ metric components as functions of $r/r_h$, for the hairy BH solution with $p= 1.1$. We show both the  solution computed solving numerically the field equations, and the semi-analytical approximation up to fourth order in terms of continued fractions.
For the numerical solution the process we follow is a shooting approach with respect to the non-GR parameter $b_1$, integrating from the horizon outwards, so that at asymptotic infinity the exponentially diverging solution vanishes. For comparison, we also show the Schwarzschild solution.

In the following we will use the semi-analytical background instead of the fully numerical one. This allows us to perform computations significantly faster, which is important considering the complexity of the perturbation equations in the gravitational sector, discussed in the next section.

\subsection{Domain of existence}\label{subsec:domain}
\begin{figure}
    \centering
    \includegraphics{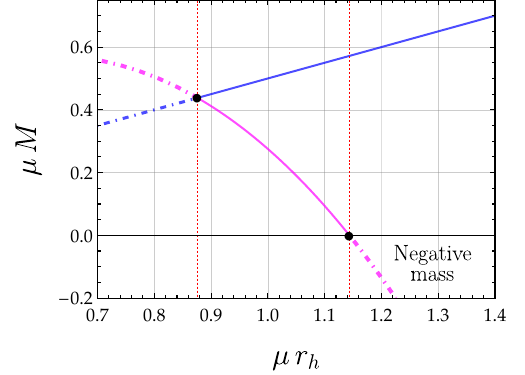}
    \caption{Domain of existence of BHs in EW gravity. The bifurcation point from the GR solution is found at $p_{\text{min}} \simeq 0.876$, while the the upper value of the positive mass range is located at $p_{\text{max}} \simeq 1.143$. The dashed part of the curves corresponds to solutions either radially unstable, or  having negative mass. }
    \label{fig:domain}
\end{figure}

The branch of non-GR BH solutions can be found for\,\cite{Kokkotas:2017zwt}
\begin{equation}
    0.876\lesssim \,p\, \lesssim 1.143\, ,
\end{equation}
where we remind the reader that the dimensionless parameter $p$ is defined as $p\equiv {r_h}/{\sqrt{2\alpha}} \equiv r_h \mu$ .
In Fig.\,\ref{fig:domain} we show the domain of existence and of radial stability of hairy BHs in EW gravity, corresponding to the decreasing curve. The vertical dashed line highlights the allowed range, which starts at the bifurcation point from the Schwarzschild branch, and ends at the point where solutions have vanishing mass.  The straight line, instead, corresponds to the solutions of the Schwarzschild branch. 

Note that hairy BHs only exist below a critical mass 
\begin{equation}
M_{c}\simeq 0.438/\mu\,, \label{eq:minmass}
\end{equation}
where Schwarzschild BHs are unstable. As variance with the Schwarzschild case, the mass-radius curve, $M(r_h)$, of the hairy BH is a decreasing function.

\section{Perturbations}
\label{sec:perturbations}
%
In the following  we shall study linear perturbations of the spacetime, which we introduce through the following expansion:
\begin{align}
g_{\mu\nu}&=\bar{g}_{\mu\nu}+\varepsilon\, \delta g_{\mu\nu}\,,\\
f_{\mu\nu}&=\bar{f}_{\mu\nu}+\varepsilon\, \delta f_{\mu\nu}\,.
\end{align}
where $\varepsilon$ is a bookkeeping parameter, $\bar{g}_{\mu\nu}$ and $\bar{f}_{\mu\nu}$ are the background metric and auxiliary field corresponding to a static, spherically symmetric BH solution, and $\delta g_{\mu\nu}$, $\delta f_{\mu\nu}$ are their respective perturbations.
Hereafter, any quantity evaluated on the non-perturbed background will be denoted with a bar. 
At first order in the perturbations, Eq.~\eqref{eq:field_g} gives:
\begin{equation}
    \delta\mathcal{E}^{(g)}_{\mu\nu}\equiv
    \delta G_{\mu\nu} 
   + \mu^2 \left(\delta f_{\mu\nu}-\bar{g}_{\mu\nu}\,\delta f
    \right)=0
    \;,
\label{eq:pert_g}
\end{equation}
while Eq.\,\eqref{eq:field_f} gives:
\begin{align}
    &\delta\mathcal{E}^{(f)}_{\mu\nu}\equiv
    \bar{f}_{\mu\nu}(\delta R - 2\mu^2\delta f)-
    \frac{1}{2}\bar{f}^{\rho\sigma}(2\bar{g}_{\mu\nu}\delta G_{\rho\sigma}+\nonumber\\
    &\mu^2 \bar{f}_{\rho\sigma}\delta g_{\mu\nu})
    +\delta (\Box f_{\mu\nu}-2\nabla_\rho\nabla_{(\mu}f_{\nu)}{}^\rho)+\nonumber\\
    &4\bar{f}_{(\mu}{}^\rho\delta G_{\nu)\rho}+4\mu^2\bar{f}_{(\mu\rho}\delta f_{\nu)}{}^\rho-\bar{R}_{\rho\sigma}\bar{g}_{\mu\nu}\delta f^{\rho\sigma}+\nonumber\\
    &4\bar{R}_{(\mu\rho}\delta f_{\nu)}{}^\rho+\delta g_{\mu\nu}\bar{\nabla}_\rho\bar{\nabla}_\sigma \bar{f}^{\sigma\rho}-\bar{R}^{\rho\sigma}\bar{f}_{\rho\sigma}\delta g_{\mu\nu}-\nonumber\\
    &\mu^2\bar{f}_{\rho\sigma}\bar{g}_{\mu\nu}\delta f^{\rho\sigma}-\delta f \bar{R}_{\mu\nu}+\bar{g}_{\mu\nu}\delta(\nabla_\rho\nabla_\sigma f^{\sigma\rho})=0\, .\label{eq:pert_f}
\end{align}
For simplicity, we shall restrict our study to perturbations with {\it axial parity}. Expanding the perturbations as in Eqs.\,\eqref{eq:pert_ansatz_g}-\eqref{eq:pert_ansatz_f}, the trace of the auxiliary field $f$ --~vanishing in the background~-- is also vanishing at linear order in the perturbations. Thus,
\begin{equation}
    \delta f=\delta R=0\,.
\end{equation}
Moreover, the linearized   Eq.\,\eqref{eq:nabla_f} gives:
\begin{align}
    \delta\mathcal{E}_{\mu}\equiv&
    \bar{\nabla}_\nu \delta f^\nu_\mu+\frac{1}{2}(\bar{f}_{\mu}^\nu\bar{\nabla}_\nu\delta g-\bar{f}^{\nu\rho}\bar{\nabla}_\rho\delta g_\nu^\rho )\nonumber\\
&    -\delta g^{\nu\rho}\bar{\nabla}_\rho \bar{f}_{\mu\nu}-\bar{f}_\mu^\nu\bar{\nabla}_\rho\delta g_\nu^\rho=0\, .
\label{eq:nd}
\end{align}

\subsection{Schwarzschild background}
\label{subsec:RT_flat}
Let us consider the case of a Schwarzschild background,  i.e. $\bar{R}_{\mu\nu}=0$, which leads to $\bar{f}_{\mu\nu} = 0$.
In this case, the perturbation equations~\eqref{eq:pert_g}-\eqref{eq:pert_f} reduce to 
\begin{align}
    &
    \delta\mathcal{E}^{(g)}_{\mu\nu}=
    \delta G_{\mu\nu} + \mu^2 \delta f_{\mu\nu}=0
    \label{eq:linear-Ricciflat-eom-g}\\
    &
    \delta\mathcal{E}^{(f)}_{\mu\nu}=
    \bar{\Box} \delta f_{\mu\nu}+2\bar{R}_{\mu\sigma\nu\rho}\delta f^{\sigma\rho}-\mu^2\delta f_{\mu\nu}=0\; .
\label{eq:linear-Ricciflat-eom-f}
\end{align}
From the above, it is evident that the massive perturbations $\delta f_{\mu\nu}$ decouple from $\delta g_{\mu\nu}$. Indeed, setting $\delta f_{\mu\nu}=0$,
Eq.\,\eqref{eq:linear-Ricciflat-eom-f} is satisfied, and Eq.\,\eqref{eq:linear-Ricciflat-eom-g}
reduces to the standard Regge-Wheeler equation\,\cite{Regge:1957td}, leading to the QNMs of Schwarzschild BHs in GR. When $\delta f_{\mu\nu}\neq0$, Eqs.\,\eqref{eq:linear-Ricciflat-eom-g}-\eqref{eq:linear-Ricciflat-eom-f} yield a set of equations for the massive spin-two  degrees of freedom, which coincide with those presented in~\cite{Brito:2013wya,OuldElHadj:2024psw} after substituting the Schwarzschild metric. Although those perturbations are massive and do propagate at infinity, they nevertheless source ordinary metric perturbations through Eq.~\eqref{eq:linear-Ricciflat-eom-g}.
Axial perturbations of the massive spin-2 excitation $\delta f_{\mu\nu}$ can be decomposed in spherical tensor harmonics (see Appendix~\ref{ap:perturbations}), leading - for each value of the harmonic indexes - to two vector perturbations, $F_0(r)$, $F_1(r)$,  and one tensor perturbation, $F_2(r)$. 
In the massless limit, the only dynamical perturbation in the decomposition of $\delta f_{\mu\nu}$ is the tensor perturbation $F_2$, while the vector ones become pure gauge\,\footnote{Note that in this limit one does not recover, for $\delta f_{\mu\nu}$, the same decomposition as the (axial) massless field $\delta g_{\mu\nu}$, because the tensorial decomposition of $\delta f_{\mu\nu}$ is different from the Regge-Wheeler gauge employed in the decomposition of $\delta g_{\mu\nu}$.
In particular, on a Schwarzschild background $\delta f_{\mu\nu}$ is traceless and transverse. 
}. 

To compute the QNMs, we first solve Eq.\,\eqref{eq:linear-Ricciflat-eom-f} (which does not depend on $\delta g_{\mu\nu}$) for $\delta f_{\mu\nu}$. Then, we replace the solution in Eq.\,\eqref{eq:linear-Ricciflat-eom-g}.
We expand the perturbations $\delta f_{\mu\nu}$, $\delta g_{\mu\nu}$ with axial parity as in Eqs.\,\eqref{eq:pert_ansatz_g},~\eqref{eq:pert_ansatz_f}, in terms of the perturbation functions  $h_a^{\ell m}(r)$ ($a=0,1$), and $F_i^{\ell m}(r)$ ($i=0,1,2$); for brevity, we do not write explicitly the harmonic indexes $(l,m)$ (note that the azimuthal number $m$ is anyway degenerate due to the spherical symmetry of the background). 
Although there exists also a dynamical dipolar ($\ell=1$) axial mode~\cite{Brito:2013wya}, the latter sources dipolar gravitational perturbations which are not dynamical. In the following we will  focus on $\ell\geq2$ modes.

The non-divergence constraint\,\eqref{eq:nd} yields
\begin{equation}
    \frac{F_1}{2}\left(\frac{A'}{A}+\frac{B'}{B}+\frac{4}{r}\right)+\frac{i \omega {F_0}}{A B}+\frac{\left(\Lambda -2\right) {F_2}}{r^2 B}+{F_1}'=0,\label{eq:constraint}
\end{equation}
while the $(t,\varphi),\,(r,\varphi)$ and $(\theta,\varphi)$  components of the perturbation equation\,\eqref{eq:linear-Ricciflat-eom-f} give 
\begin{align}
    &F_0''+\frac{F_0'}{2} \left(\frac{B'}{B}-\frac{A'}{A}\right)-\frac{i F_1 \omega A'}{A}+\frac{F_0}{4 A^2 B r^2}\times\nonumber\\
    &\big[B r^2 A'^2-2 A^2 \left[r B'+2 B+2 \left(\mu ^2 r^2+\Lambda -1\right)\right]+\label{eq:t_varphi}\nonumber\\
    &A r \left(r \left(4 \omega^2-A' B'\right)+2 B \left(A'-r A''\right)\right)\big]=0\,,\\[5mm]
    &F_1''+\frac{F_1'}{2} \left(\frac{A'}{A}+\frac{3 B'}{B}\right)-\frac{i F_0 \omega A'}{A^2 B}-\frac{2 F_2 \left(\Lambda -2\right)}{B r^3}\nonumber\\
    &+\frac{F_1}{4 A^2 B r^2}\big[A^2 (2 r^2 B''+2 r B'-20 B-4 \mu ^2 r^2\label{eq:r_varphi}\nonumber\\
    &-4 \Lambda +4)+A r \left(r A' B'-2 B A'+4 r w^2\right)\nonumber\\
    &-B r^2 A'^2\big]=0\,,\\[5mm]
    &F_2''+\frac{F_2'}{2} \left(\frac{A'}{A}+\frac{B'}{B}-\frac{4}{r}\right)-\frac{2 F_1}{r}+\frac{F_2}{A B r^2}\times\label{eq:theta_varphi}\nonumber\\
    &\big[r \left(r \omega^2-B A'\right)-A \left(r B'-2 B+\mu ^2 r^2+\Lambda -2\right)\big]\nonumber\\
    &=0\,,
\end{align}
where $\Lambda\equiv\ell(\ell+1)$.
Finally, from Eq.~\eqref{eq:linear-Ricciflat-eom-g}, by eliminating $h_0$, we find
\begin{align}
    &h_1''+h_1' \bigg(\frac{3 A'}{2 A}+\frac{3 B'}{2 B}-\frac{2}{r}\bigg)-\frac{2 \mu ^2 F_2 (r A'-2 A)}{r A B}\nonumber\\
    &+\frac{1}{2 r^2 A B}\big[h_1 (r^2 (B A''+2 A' B'+2 w^2)+A (r^2 B''\nonumber\\
    &+4 B-2 \Lambda))\big]-\frac{2 \mu ^2 F_1}{B}-\frac{2 \mu ^2F_2'}{B}=0\,.\label{eq:h1}
\end{align}
We may now eliminate $F_0$ by making use of~\eqref{eq:constraint}, in which case the decoupled equations for the massive perturbations can be written as a system of Schr\"odinger-type equations
\begin{equation}
    \frac{d^2}{dr_*^2}
    \begin{pmatrix}
        Q\\
        Z
    \end{pmatrix}
    +
    \begin{pmatrix}
        \hat{V}_{11} & \hat{V}_{12}\\
        \hat{V}_{21} & \hat{V}_{22}
    \end{pmatrix}
    \begin{pmatrix}
        Q\\
        Z
    \end{pmatrix}=0\, ,
    \label{eq:Ricci_tensor_flat_system}
\end{equation}
where $Q\equiv F_1\sqrt{A B} , \; Z\equiv F_2/r$ and the matrix elements are given by
\begin{align}
    \hat{V}_{11}=&-A\mu^2+\big[-2 A^2 \left(-r B'+10 B+2 \left(\Lambda -1\right)\right)\nonumber\\
    &+B r^2 A'^2+A r (-2 B r A''-r A' B'+6 B A'\nonumber\\
    &+4 r \omega^2)\big](4r^2 A)^{-1}\,,\\[2mm]
    \hat{V}_{12}=&-r^{-2}\left(\Lambda -2\right) \sqrt{A B}\left(2 A-r A'\right)\,,\\[2mm]
    \hat{V}_{21}=&-2r^{-2}\sqrt{A B}\, ,\\[2mm]
    \hat{V}_{22}=&-A\mu^2+\big[r (2 r \omega^2-B A')-A (r B'\nonumber\\
    &+2 (\mu ^2 r^2+\Lambda-2))(2 r^2)^{-1}\big]\,.
\end{align}
The terms $-A\mu^2$ in $\hat{V}_{11}$ and $\hat{V}_{22}$ are an indication  that $Q$ and $Z$ correspond to massive modes propagating with mass $\mu^2$.
Notice that with these redefinitions for $F_1$ and $F_2$ we cannot simultaneously describe all three perturbation functions $(F_1,F_2,h_1)$ in a Schr\"odinger-type system due to the presence of the $F_2'$ term in~\eqref{eq:h1}. 

In order to study the full QNM spectrum, we solved the following system describing both massless and massive degrees of freedom:
\begin{equation}
    \frac{d^2}{dr^2}\boldsymbol{\Psi}+\boldsymbol{P}\frac{d}{dr}\boldsymbol{\Psi}+\boldsymbol{V}\Psi=0
\end{equation}
where $\boldsymbol{\Psi}=\big(\Psi^{(1)},\Psi^{(2)},\Psi^{(3)}\big)\equiv (h_1,F_1,F_2)$, and
\begin{equation}
    \boldsymbol{P}=
    \begin{pmatrix}
        {P}_{11} & 0 & {P}_{13}\\
        0 & {P}_{22} & 0\\
        0 & 0 & {P}_{33}
    \end{pmatrix}\, , \;
    \boldsymbol{V}=
    \begin{pmatrix}
        {V}_{11} & {V}_{23} & {V}_{13}\\
        0 & {V}_{22} & {V}_{23}\\
        0 & {V}_{32} & {V}_{33}
    \end{pmatrix} \, ,
\end{equation}
where the matrix elements of $\boldsymbol{P}$ and $\boldsymbol{V}$ are functions of the background solution.
In order to solve this system of equations numerically we need to impose appropriate boundary conditions at the horizon and at infinity.
\begin{figure}[t]
    \centering
    \includegraphics[width=\linewidth]{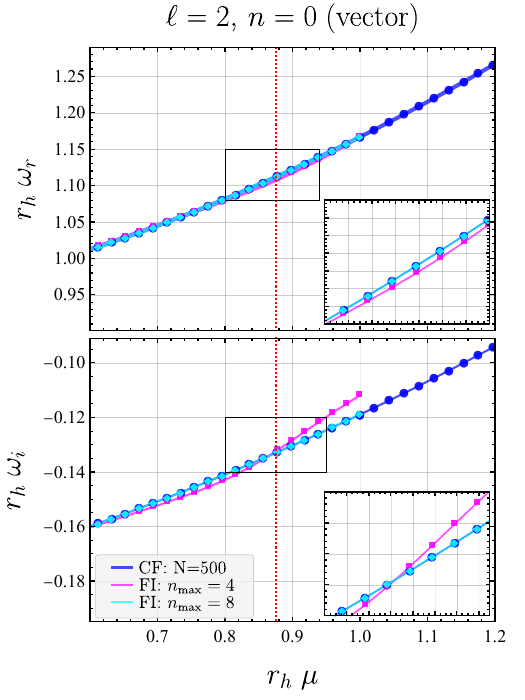}
    \caption{Real (upper panel) and imaginary (lower panel) parts of the fundamental axial, $\ell=2$ massive vector mode, normalized with $r_h$, as a function of $p=r_h\mu$, in the case of Schwarzschild background.
    We show the results obtained with the DI method (with $n_{\rm max}=4,8$) and with the CF method (performed with $N=500$ steps). The inset shows a detail of the comparison between different methods. The Schwarzschild background is stable for $p$ larger than the threshold value denoted by the vertical dotted line.}
\label{fig:Ricci_tensor_flat_1}
\end{figure}
\begin{figure}[t]
    \centering
    \includegraphics[width=\linewidth]{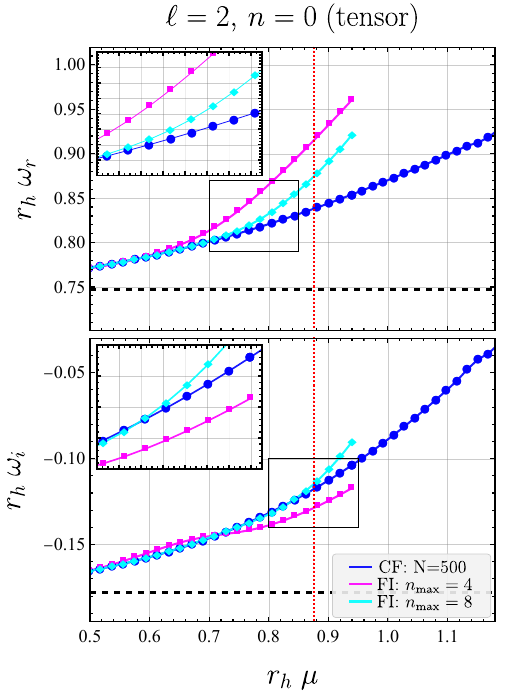}
    \caption{Same as Fig.~\ref{fig:Ricci_tensor_flat_1}, for tensor modes. The horizontal dashed line corresponds to the massless mode, which coincides with that of GR BHs.}
\label{fig:Ricci_tensor_flat_2}
\end{figure}

For ingoing waves at the horizon, ${\Psi}^{(j)}(r)\sim e^{-i\omega r_*}$.
Therefore, we expand
\begin{align}
    h_1(r)=&\;e^{-i\omega r}(r-2M)^{-2iM\omega}\sum_{n=0} \tilde{h}_1^{(n)}(r-2M)^{n-1}\,,\label{eq:b_hor_1}\\
    F_1(r)=&\;e^{-i\omega r}(r-2M)^{-2iM\omega}\sum_{n=0} f_1^{(n)}(r-2M)^{n-1}\,, \label{eq:b_hor_2}\\
    F_2(r)=&\;e^{-i\omega r}(r-2M)^{-2iM\omega}\sum_{n=0} f_2^{(n)}(r-2M)^{n}\,.\label{eq:b_hor_3}
\end{align}
By substituting~\eqref{eq:b_hor_1}-\eqref{eq:b_hor_2} into~\eqref{eq:Ricci_tensor_flat_system} we solve for $f_{1,2}^{(n)}$ with $n\ge 1$ in terms of $f_{1,2}^{(0)}$.
Substitution of~\eqref{eq:b_hor_3} in~\eqref{eq:pert_g} yields another free parameter, namely $\tilde{h}_1^{(0)}$. Therefore, we have three free parameters in total at the horizon, namely $(h_1^{(0)},f_{1}^{(0)},f_{2}^{(0)})$.

For QNMs, we impose outgoing wave boundary condition at $r\to\infty$. This leads to:
\begin{align}
    h_1(r)=&\;e^{ikr}r^x\sum_{n=0} \frac{H_{1a}^{(n)}}{r^{n-1}}+e^{i\omega r}r^{2iM\omega}\sum_{n=0} \frac{H_{1b}^{(n)}}{r^{n-1}}\,,\label{eq:b_inf_1}\\
    F_1(r)=&\;e^{ikr}r^x\sum_{n=0} \frac{F_1^{(n)}}{r^{n}}\,,\label{eq:b_inf_2}\\
    F_2(r)=&\;e^{ikr}r^x\sum_{n=0} \frac{F_2^{(n)}}{r^{n-1}}\,,\label{eq:b_inf_3}
\end{align}
where $k=\sqrt{\omega^2-\mu^2}$ and $x=M(\mu^2-2\omega^2)/(ik)$. Since we are studying QNMs, we look for solutions with $\omega>\mu$; we note that the equation for $\delta f_{\mu\nu}$ also admits quasi-bound state solutions with $\omega<\mu$~\cite{Brito:2013wya}.

We also note that the perturbation function $h_1$ has two contributions at asymptotic infinity (Eq.\,\eqref{eq:b_inf_3}), corresponding to the  massless and massive degrees of freedom. This is a consequence of the fact that $\delta g_{\mu\nu}$ is coupled with $\delta f_{\mu\nu}$ through\,\eqref{eq:field_g}.
By substituting Eqs.\,\eqref{eq:b_inf_1}-\eqref{eq:b_inf_3} in Eqs.\,\eqref{eq:field_g}-\eqref{eq:field_f} we find the coefficients $F_{1,2}^{(n)}$ with $n\ge 1$ in terms of $F_{1,2}^{(0)}$, while $H_{1a}^{(n)}$ is determined in terms of $(F_1^{(0)},F_2^{(0)})$ for $n\ge 0$. Additionally, $H_{1b}^{(n)}$ is found in terms of $H_{1b}^{(0)}$ for $n\ge 1$. Therefore, we have three free parameters at infinity, namely $(H_{1b}^{(0)},F_1^{(0)},F_2^{(0)})$.

The forward direct integration (DI) method we use here is presented in Appendix~\ref{ap:methods}.
The asymptotic expansion at the horizon, Eqs.\,\eqref{eq:b_hor_1}-\eqref{eq:b_hor_3}, is performed up to $n=4$. Our convergence tests show that increasing the order of this expansion does not significantly affect the accuracy of our results, leading to truncation errors smaller than $1\%$.
The asymptotic expansion at infinity, Eqs.\,\eqref{eq:b_inf_1}-\eqref{eq:b_inf_3}, is performed up to $n=n_{\rm max}$. The value of $n_{\rm max}$ required to have a good accuracy of the results depends on the value of the mass of the spin-2 mode, and of the BH mass: larger values of  $\mu r_h$ require more terms in the asymptotic expansion at infinity. 

Since we are considering the case of Schwarzschild background, it is also possible to employ a continued fraction (CF) approach, which does not have the numerical problems of the DI method\,\cite{Chandrasekhar:1975zza,Leaver:1985ax}.
We found that considering $N=500$  steps in the CF approach is sufficient, since increasing that number by even an order of magnitude does not affect the results significantly.
For each $\ell\geq2$, we find two independent  classes of axial QNMs, corresponding to vector and tensor modes of the massive spin-2 field~\cite{Brito:2013wya}. 

Summarizing, for each value of the BH mass (i.e. of the horizon radius), for each $\ell$, we find three series of modes with axial parity, parametrized as customarily by the principal number $n$: (i) the massless modes, which coincide with those of GR; (ii) the massive vector modes; (ii) the massive tensor modes. 

We present our results in Figs.\,\ref{fig:Ricci_tensor_flat_1} \ref{fig:Ricci_tensor_flat_2}, where we show the fundamental ($n=0$) $\ell=2$ massive vector and tensor modes, respectively, computed using the DI approach with $n_{\text{max}}=4,8$, and the CF approach. We find that the DI results, increasing $n_{max}$, converge towards the CF results, which we use as a benchmark. 
The vector mode computed using the DI approach with $n_{\text{max}}=4$ is a very good approximation, within $1\%$, of the CF result  for $p\lesssim0.6$, while for larger values of $p$ we need $n_{\max}=8$ to reach the same level of accuracy. {\bf For the tensor mode, the DI approach with $n_{\text{max}}=4$ is accurate within $1\%$ for $p\lesssim0.6$, while that with  $n_{\text{max}}=8$ has the same accuracy for $p\lesssim0.75$. For larger $p$ the error increases but is 
smaller than
$5\%$ for $p\lesssim 0.95$.}
In Fig.~\ref{fig:Ricci_tensor_flat_2} we also show the fundamental massless mode, which also belongs to the spectrum, and is the same as in GR. It coincides with the $\mu\to0$ limit of the massive tensor mode, while we remind that the massive vector mode is not dynamical in this limit (see Appendix\,\ref{ap:perturbations}).

We remark that in Figs.\,\ref{fig:Ricci_tensor_flat_1}, \ref{fig:Ricci_tensor_flat_2}, the Schwarzschild background is stable for $p$ larger than the threshold value denoted by the vertical dotted line. The unstable region is shown to clarify how the agreement between DI and CF approaches improves by increasing $n_{\rm max}$. 

\subsection{Hairy BH background}
\label{subsec:R_flat}
%
In this case the analysis is more complicated since we have to solve the full system of coupled equations\,\eqref{eq:pert_g}, \eqref{eq:pert_f}.
As we have shown, static, spherically symmetric, and asymptotically flat BHs in quadratic gravity are Ricci-scalar flat. Therefore $\bar{R}=\bar{f}=0$ and ~\eqref{eq:nabla_f} yields $\bar{\nabla}^{\nu}\bar{f}_{\mu\nu}=0$.
The constraint equation is $\nabla^\mu f_{\mu\nu}=0$.

From~\eqref{eq:pert_g} we have three nontrivial components. Below we show their schematic  form (note that the coefficients depend on the background coordinates and on $\omega$):
\begin{align}
    \begin{split}
        (t\varphi):\;
        & \; c_1^{(t\varphi)} h_0''+c_2^{(t\varphi)} h_0'+c_3^{(t\varphi)} h_1'+c_4^{(t\varphi)} h_0\\
        & +c_5^{(t\varphi)} h_1+c_6^{(t\varphi)} F_0=0\,,
    \end{split}
    \\[2mm]
    (r\varphi):\;
    & c_1^{(r\varphi)} h_0'+c_2^{(r\varphi)} h_0+c_3^{(r\varphi)} h_1+c_4^{(r\varphi)} F_1=0\,,
    \\[2mm]
    (\theta\varphi):\;
    & c_1^{(\theta\varphi)} h_1'+c_2^{(\theta\varphi)} h_0+c_3^{(\theta\varphi)} h_1+c_4^{(\theta\varphi)} F_2=0\,.
\end{align}
From~\eqref{eq:pert_f}
\begin{align}
    \begin{split}
        (t\varphi):\;
        & \;d_1^{(t\varphi)} F_0''+d_2^{(t\varphi)} h_0''+d_3^{(t\varphi)} F_0'+d_4^{(t\varphi)} F_1'\\
        & +d_5^{(t\varphi)} h_0'+d_6^{(t\varphi)} h_1'+d_7^{(t\varphi)} F_0+d_8^{(t\varphi)} F_1\\
        & +d_9^{(t\varphi)} F_2+d_{11}^{(t\varphi)} h_0+d_{12}^{(t\varphi)} h_1=0\,,
    \end{split}
    \\[2mm]
    \begin{split}
        (r\varphi):\;
        & \; d_1^{(r\varphi)} F_0'+d_2^{(r\varphi)} F_2'+d_3^{(r\varphi)} h_0'+d_4^{(r\varphi)} F_0\\
        & +d_5^{(r\varphi)} F_1+d_6^{(r\varphi)} F_2+d_7^{(r\varphi)} h_0\\
        & +d_8^{(r\varphi)} h_1=0\,,
    \end{split}
    \\[2mm]
    \begin{split}
        (\theta\varphi):\;
        & \; d_1^{(\theta\varphi)} F_2''+d_2^{(\theta\varphi)} F_1'+d_3^{(\theta\varphi)} F_2'+d_4^{(\theta\varphi)} h_1'\\
        & +d_5^{(\theta\varphi)} F_0+d_6^{(\theta\varphi)} F_1+d_7^{(\theta\varphi)} F_2+d_8^{(\theta\varphi)} h_0\\
        & +d_9^{(\theta\varphi)} h_1=0\,.
    \end{split}
\end{align}
From Eq.~\eqref{eq:nd} we have
\begin{equation}
    p_1 F_1'+p_2 h_1'+p_3 F_0+p_4 F_1+p_5 F_2+p_6 h_0+p_7 h_1=0\,.
\end{equation}
We then solve the constraint equation together with the $(\theta\varphi)$ component of\,\eqref{eq:pert_g} for $h_0,\,F_0$ and we substitute those into the $(r\varphi)$ component of\,\eqref{eq:pert_g} and the $(r\varphi)$ and $(\theta\varphi)$ components of\,\eqref{eq:pert_f}. We end up with the following coupled system of equations for the three perturbation functions $h_1,\,F_1,\,F_2$:
\begin{equation}
    \frac{d^2}{dr^2}\boldsymbol{\Psi}+\boldsymbol{P^h}\frac{d}{dr}\boldsymbol{\Psi}+\boldsymbol{V^h}\Psi=0
\end{equation}
where $\boldsymbol{\Psi}=(\Psi^{(1)},\Psi^{(2)},\Psi^{(3)})\equiv (h_1,F_1,F_2)$, 
\begin{equation}
    \boldsymbol{P^h}=
    \begin{pmatrix}
        {P}^h_{11} & 0 & {P}^h_{13}\\
        {P}^h_{21} & {P}^h_{22} & {P}^h_{23}\\
        {P}^h_{31} & 0 & {P}^h_{33}
    \end{pmatrix}\,, \;
    \boldsymbol{V^h}=
    \begin{pmatrix}
        {V}^h_{11} & {V}^h_{12} & {V}^h_{13}\\
        {V}^h_{21} & {V}^h_{22} & {V}^h_{23}\\
        {V}^h_{31} & {V}^h_{23} & {V}^h_{33}
    \end{pmatrix} \,,
\label{eq:axial_system}
\end{equation}
where the matrix elements depend on the hairy background solutions, and the overscript $h$ stands for the hairy background.
At the horizon, assuming ingoing wave boundary conditions, we find
\begin{align}
    h_1(r)=&\;e^{-i\omega r}(r-r_h)^{-i\omega/\sqrt{c\,b_1}}\sum_{n=0} h_1^{(n)}(r-r_h)^{n-1}\,,\\
    F_1(r)=&\;e^{-i\omega r}(r-r_h)^{-i\omega/\sqrt{c\,b_1}}\sum_{n=0} f_1^{(n)}(r-r_h)^{n-1}\,,\\
    F_2(r)=&\;e^{-i\omega r}(r-r_h)^{-i\omega/\sqrt{c\,b_1}}\sum_{n=0} f_2^{(n)}(r-r_h)^{n}\,,
\end{align}
where $c$ and $b_1$ are the coefficients appearing in the near-horizon expansions of the background solutions~\eqref{eq:bg_horizon_1},~\eqref{eq:bg_horizon_2}.
By substituting in~\eqref{eq:axial_system} we can solve for $h_1^{(n)},\,F_{1,2}^{(n)}$ with $n\ge 1$ in terms of three free parameters, namely $(F_1^{(0)},F_2^{(0)},h_1^{(0)})$.
\begin{figure}[t]
    \centering
    \includegraphics[width=\linewidth]{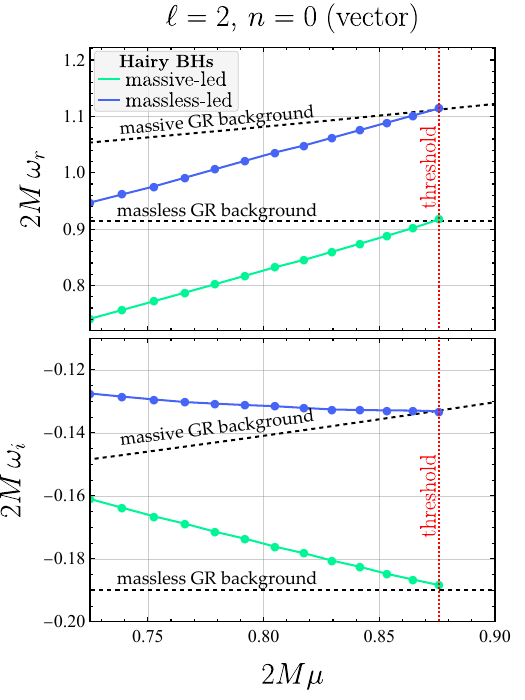}
    \caption{Real (upper panel) and imaginary (lower panel) parts of the fundamental axial, $\ell=2$ vector modes, as functions of $2M\mu$ in the case of hairy BH background. The solid lines correspond to the frequencies of the massless-led and massive-led modes, computed using the DI approach with $n_{\text{max}}=8$. For comparison, we also show the frequencies of the corresponding modes in the case of Schwarzschild background (dashed lines). The  vertical dashed line denotes the threshold $M=M_c$.}
\label{fig:Ricci_scalar_flat_1}
\end{figure}
\begin{figure}[t]
    \centering
    \includegraphics[width=\linewidth]{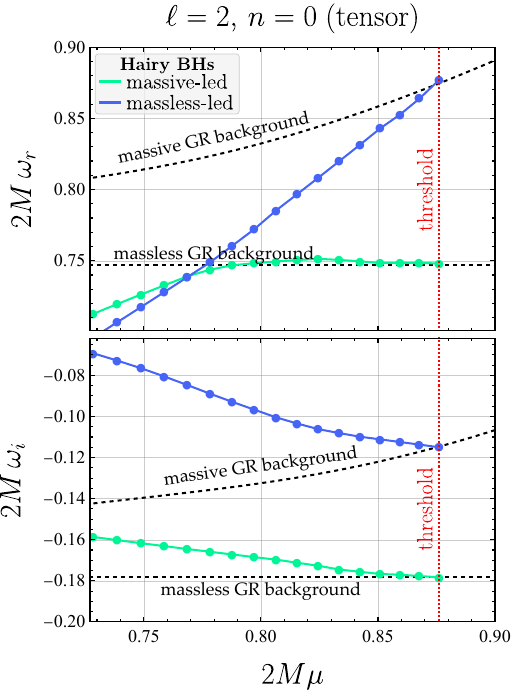}
    \caption{Same as Fig.~\ref{fig:Ricci_scalar_flat_1} for tensor modes.}
\label{fig:Ricci_scalar_flat_2}
\end{figure}

For the forward integration, the expansions at infinity have to be modified in order to incorporate the contribution from the massive and massless modes as well. For QNMs, we look for solutions with $\omega>\mu$ and impose outgoing boundary conditions at infinity, leading to:
\begin{align}
    h_1(r)=&\;e^{ik r}r^x\sum_{n=0} \frac{H_{1a}^{(n)}}{r^{n-1}}+e^{i\omega r}r^{2iM\omega}\sum_{n=0} \frac{H_{1b}^{(n)}}{r^{n-1}}\,, \label{eq:expasions_infinity_1}\\
    F_1(r)=&\;e^{ik r}r^x\sum_{n=0} \frac{F_{1a}^{(n)}}{r^{n}}+e^{i\omega r}r^{2iM\omega}\sum_{n=0} \frac{F_{1b}^{(n)}}{r^{n}}\,, \label{eq:expasions_infinity_2}\\
    F_2(r)=&\;e^{ik r}r^x\sum_{n=0} \frac{F_{2a}^{(n)}}{r^{n-1}}+e^{i\omega r}r^{2iM\omega}\sum_{n=0} \frac{F_{2b}^{(n)}}{r^{n-1}}\,, \label{eq:expasions_infinity_3}
 \end{align}
where $k=\sqrt{\omega^2-\mu^2}$ and $x=M(\mu^2-2\omega^2)/(ik)$.
After substitution in~\eqref{eq:axial_system} and solving order by order we determine all coefficients in terms of three parameters, namely $(H_{1b}^{(0)},F_{1a}^{(0)},F_{2a}^{(0)})$.
To find the QNMs we perform three forward integrations by fixing the horizon parameters $(h_1^{(0)},F_{1}^{(0)},F_{1}^{(0)})$, and then demand that the coefficients associated with the ingoing waves at infinity vanish, according to the method explained in Appendix~\ref{ap:methods}.

Since the massless and massive perturbations are coupled, we do not have purely massive and purely massless QNM solutions. We find, instead, {\it massless-led} solutions, in which the contribution of the massive field is due to the coupling in Eq.\,\eqref{eq:field_g}, and similarly {\it massive-led solutions}. Therefore, we have four series of modes: the massless-led vector and tensor modes (which in this case are different from those of GR), and the massive-led vector and tensor modes.

We present our results for the $\ell=2$, $n=0$  vector and tensor modes in Figs.\,\ref{fig:Ricci_scalar_flat_1} and~\ref{fig:Ricci_scalar_flat_2}, respectively.
The solid lines respresent the frequencies of the massive-led and massless-led modes, as functions of $2M \mu$. For comparison, we also show the frequencies of the corresponding modes in the case of the Schwarzschild background (dashed line). Finally, the vertical dashed line denotes the critical mass $M_c$.
\begin{figure}[t]
    \centering
    \includegraphics[width=0.98\linewidth]{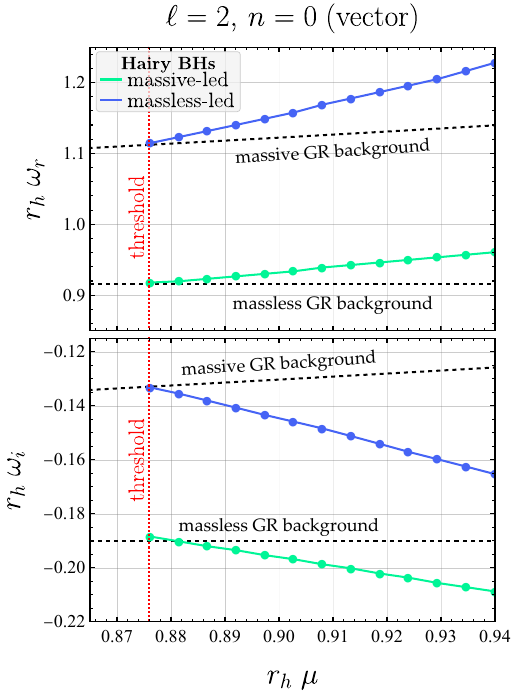}
    \caption{Same as Fig.~\ref{fig:Ricci_scalar_flat_1} but with a different horizontal and vertical scaling. Notice that the non-GR QNMs bifurcate towards the opposite direction with respect to Fig.~\ref{fig:Ricci_scalar_flat_1}. This is a consequence of the fact that for the hairy solutions increasing the BH mass corresponds to smaller horizon radii.}
\label{fig:Ricci_scalar_flat_3}
\end{figure}
\begin{figure}[t]
    \centering
    \includegraphics[width=0.98\linewidth]{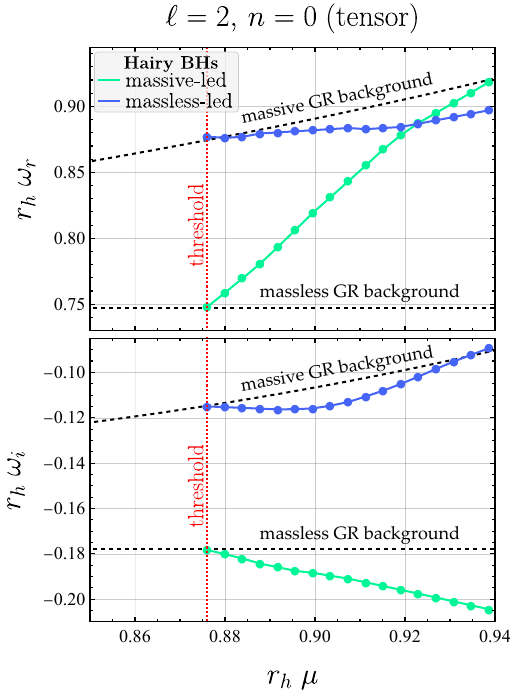}
    \caption{Same as Fig.~\ref{fig:Ricci_scalar_flat_3} for tensor modes.}
\label{fig:Ricci_scalar_flat_4}
\end{figure}

For the vector modes showed in Fig.\,\ref{fig:Ricci_scalar_flat_1} we notice that within the parameter space we explore, both the decay rate and oscillation frequency increase as the deviation from GR increases, \textit{i.e.} for smaller values of $\mu$ (larger values of the coupling constant $\alpha$) at a fixed BH mass $M$.
In the case of the tensor modes depicted in Fig.\,\ref{fig:Ricci_scalar_flat_2} the behavior of the massive-led mode is non-monotonic and harder to describe.
For clarity, in Figs.\,\ref{fig:Ricci_scalar_flat_3} and~\ref{fig:Ricci_scalar_flat_4}, we  present the same QNM modes as functions of $p=\mu r_h$, and  normalizing the frequencies and the mass of the spin-2 mode with the BH  horizon radius rather than with its mass.

At variance with with the case of the Schwarzschild background discussed in the previous subsection, the hairy background is numerical or semi-analytical, and therefore  a CF approach for the calculation of QNMs is not as straightforwardly applicable. Thus, we only employ the DI approach.
Due to the limitations of the DI method, 
we do expect  numerical errors to be present in this computation. However, the analysis of the Schwarzschild background case gives strong indication that the DI approach with $n_{\text{max}}=8$ is accurate enough to give the QNM frequencies within a few percent error for the mass range we are interested in.

In any case, our analysis does not aim to make highly accurate predictions, but rather to find the general structure of the ringdown spectrum of BHs in quadratic gravity, estimating the frequencies and damping times of the QNMs.

Our analysis also has the objective of probing the stability of the stationary BH solutions. To this aim, we searched for unstable modes ($\omega_i>0$) and found none both in the Schwarzschild and in the hairy BH background. This provides a strong indication of stability of the stationary BH solutions under non-radial perturbations.

\section{Phenomenology}
\label{sec:pheno}
%
In this section we discuss, in the light of our results, the phenomenology of BHs in quadratic gravity, i.e. of EW gravity (since, as discussed in Sec.\,\ref{sec:framework}, the $\beta R^2$ term in the action  affects neither the BH background nor its perturbations).

In the region where both Schwarzschild and hairy BHs are radially stable, our analysis did not find any numerical evidence for the instability of either solutions in the axial sector. We therefore assume that hairy BH solutions are stable in the range $0.876\lesssim \,p\, \lesssim 1.143$\,\cite{Kokkotas:2017zwt}, and that Schwarzschild BH solutions are stable for  $p\gtrsim 0.876$.
%

%
\subsection{Horizon radius}\label{subsec:hor}
The horizon radius of a static BH as a function of its mass, $r_h(M)$, in EW gravity coincides with that in GR, $r_h=2M$, for $M>M_c\simeq 0.438/\mu$ (Eq.\,\eqref{eq:minmass}), while it has two branches for $M<M_c$: the (unstable) Schwarzschild branch, and the hairy BH branch, in which $r_h>2M$. 
Therefore, as long as static (or stationary, see\,\cite{Sajadi:2023smm}) BHs are concerned, the deviations from GR are only visible for BHs with masses smaller than the critical mass $M_c$. 

\begin{figure}[t]
    \centering
    \includegraphics[width=\linewidth]{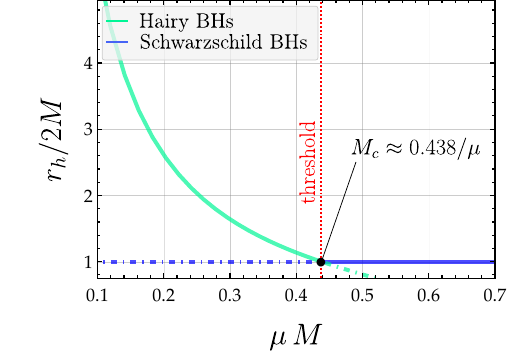}
    \caption{BH compactness as function of the BH mass, normalized with $\mu$, in GR and in EW gravity. The vertical dotted line correspond to the critical mass $M_c$. The other dashed lines correspond, to radially unstable branches. Note that for $M>M_c$, EW gravity and GR share the same (radially stable) Schwarzschild solution.}
    \label{fig:Mass-radius}
\end{figure}

In Fig.\,\ref{fig:Mass-radius} we show the BH compactness, i.e. the ratio $r_h/2M$, as a function of $M\mu=M/\sqrt{2\alpha}$, i.e. of the BH mass normalized with the coupling constant. We can see that when $M$ is significantly smaller than $M_c$, the horizon radius is much larger than the GR value. The same applies for stationary, rotating BHs\,\cite{Sajadi:2023smm}. Moreover, as noted in\,\cite{Kokkotas:2017zwt}, the decrease of the innermost stable circular orbit (ISCO) frequency is even larger than what could be inferred from the increase of the horizon radius: the product $\Omega_{\rm ISCO}M$ is smaller than the value $1/(6\sqrt{6})$ predicted by GR.

This indicates that the astrophysical BHs observed through their electromagnetic and GW emission have masses larger than the critical mass $M_c$ (in principle, the lightest observed BH may have a mass smaller than $M_c$ as long as it is very close to that value).
Indeed, the ISCO frequency characterizes the $X$-ray emission from accretion disks around BHs\,\cite{Abramowicz:2011xu}, and the frequency of the last stable orbit of a BH binary - related to the ISCO frequency - can be directly constrained from the gravitational waveforms observed in BH binary coalescences, since it cannot be larger than the observed merger frequency\,\cite{LIGOScientific:2016aoc}.
Therefore, unless lighter BHs are observed in the future, it is unlikely that observations of the stationary BH spacetime will probe quadratic gravity.

Current GW observations suggest that the lightest BH candidate observed may have masses in the range $M_{\min}\sim(2.5-4.5)\,M_\odot$\,\cite{LIGOScientific:2024elc}. This sets a bound on the coupling constant of quadratic gravity of $\sqrt\alpha\lesssim (1.1-2.0)$ km.

\subsection{BH spectroscopy \& extra modes}\label{subsec:spectr}
%
BH spectroscopy, the study of BHs from the observation of their QNMs of oscillation measured from the GW signal emitted in the ringdown stage of a binary BH coalescence\,\cite{Berti:2018vdi}, is a promising way to probe GR deviations such as those predicted by quadratic gravity.
Although the accuracy of present measurements of QNMs is probably not sufficient to provide strong constraints\,\cite{LIGOScientific:2021sio,Baibhav:2023clw,ringdownreview}, the situation will improve with future observations, especially with future detectors~\cite{Berti:2016lat,Bhagwat:2021kwv,Bhagwat:2023jwv}.
In particular, third-generation ground-based detectors such as the Einstein Telescope~~\cite{Punturo:2010zz,Maggiore:2019uih,Reitze:2019iox, Kalogera:2021bya, Hild:2010id,Branchesi:2023mws} and Cosmic Explorer~\cite{LIGOScientific:2016wof,Essick:2017wyl,Evans:2023euw} are expected to measure $\mathcal{O} (10^3)$ ringdowns/yr with $\mathcal{O}(10)\%$ precision for deviations of the subdominant frequencies from the GR predictions, whereas $\mathcal{O}(10)$ ringdowns/yr are expected to have $\mathcal{O}(1)\%$ precision.
The future space-based LISA detector\,\cite{Colpi:2024xhw} will measure at least three independent QNM parameters within $1\%$ error in ${\cal O}(100)$ events (actual numbers depend on the massive BH formation scenarios)~\cite{Bhagwat:2021kwv}.

In the case of quadratic gravity, as discussed in Sec.\,\ref{sec:perturbations}:
\begin{itemize}
\item[(i)] the QNMs of GR are modified, in the case of hairy BH background;
\item[(ii)] new classes of QNMs are present in the spectrum, associated to the new massive degrees of freedom, in both cases of Schwarzschild background and of hairy BH background.
\end{itemize}
Although we limited our study to the axial sector, we expect similar deviations of the QNMs with respect to the GR prediction in the polar sector as well.

It is also worth noting that the massless modes in the Schwarzschild background coincide with those of GR, and all other modes shown in Section\,\ref{sec:perturbations} do not converge to GR modes as $\alpha\to0$. Thus, from an EFT perspective there are no corrections to the QNM spectrum at the leading order in $\alpha$, but the new non-GR modes might be excited with amplitudes that vanish in the $\alpha\to0$ limit.

Concerning shifts in the GR QNMs, in the range shown in Figs.~\ref{fig:Ricci_scalar_flat_1}-\ref{fig:Ricci_scalar_flat_4}, deviations with respect to the ordinary massless GR mode can be as large as a few percent and thus in principle measurable by third-generation detectors. However,  as discussed in Sec.\,\ref{subsec:hor}, these shifts are only expected  if the final BH of the coalescence has $M<M_c$, and this is disfavored since the modified background would significantly affect the inspiral waveform.

A more promising possibility it that of finding new QNMs in the  ringdown of a BH with $M>M_c$, i.e. in the case of Schwarzschild background.
In this case (see Figs.~\ref{fig:Ricci_tensor_flat_1} and \ref{fig:Ricci_tensor_flat_2}) in addition to the ordinary QNMs of Schwarzschild, the spectrum contains also the modes of a massive spin-2 field~\cite{Brito:2013wya}.

The existence of these extra QNMs is by itself not sufficient to guarantee that they will be excited in the ringdown, since the latter involves ordinary gravitational perturbations.
Therefore, one needs to check if the massive spin-2 modes do affect not only the auxiliary field $f_{\mu\nu}$, but also the metric $g_{\mu\nu}$\,\cite{Crescimbeni:2024sam}. 

Let us consider the contribution of a stress-energy tensor in EW gravity, i.e. the following action
\begin{equation}
   \mathcal{I}=\frac{1}{16\pi}\, \mathcal{I}_0[g_{\mu\nu},f_{\mu\nu}]+\mathcal{I}_m [g_{\mu\nu},\psi]\, ,
\end{equation}
where $\mathcal{I}_m=\int {\rm d}^4 x \sqrt{-g} \,\mathcal{L}_m$ is the matter action depending on the metric and (schematically) on the matter fields $\psi$, while $\mathcal{I}_0$ is the EW Lagrangian presented in~\eqref{eq:Lagrangian}. In this case the right-hand side of Eq.~\eqref{eq:field_f} is modified as follows
\begin{align}
\mathcal{E}^{(f)}_{\mu\nu}=8\pi\, T_{\mu\nu}\, ,
\end{align}
where $T_{\mu\nu}\equiv -(2/\sqrt{-g})\,\delta (\sqrt{-g}\mathcal{L}_m) /\delta g^{\mu\nu}$.
This shows that ordinary matter directly sources the auxiliary field $f_{\mu\nu}$ 
rather than the metric $g_{\mu\nu}$. 
However, the latter is coupled to $f_{\mu\nu}$ through the equations of motion.

Focusing on the most relevant Schwarzschild case, the perturbation equations~\eqref{eq:linear-Ricciflat-eom-g}--\eqref{eq:linear-Ricciflat-eom-f} in the presence of matter read \cite{follow-up}
\begin{align}
    &
    \delta G_{\mu\nu} + \mu^2 \delta f_{\mu\nu}=\frac{8\pi}{3}g_{\mu\nu}T
    \label{eq:linear-Ricciflat-eom-g_matter}\,,\\
    &
    \bar{\Box}\delta f_{\mu\nu}+2\bar{R}_{\rho\mu\sigma\nu}\delta f^{\rho\sigma}
    -\mu^2 \delta f_{\mu\nu}=\nonumber\\
    & \qquad\qquad 8\pi\left(T_{\mu\nu}-\frac{1}{3}g_{\mu\nu}T +\frac{1}{3\mu^2}\bar{\nabla}_\mu\bar{\nabla}_\mu T\right)\; .
\label{eq:linear-Ricciflat-eom-fmatter}
\end{align}
Thus, matter perturbations (e.g., the merger initial conditions) would source $\delta f_{\mu\nu}$, whose spectrum consists of massive spin-2 QNMs, and that in turn sources $\delta g_{\mu\nu}$ through Eq.~\eqref{eq:linear-Ricciflat-eom-g_matter}.

As a back-of-the-envelope estimate, we can assume that $\bar\Box\sim 1/M^2$ and $\bar{R}_{\mu\sigma\nu\rho}\sim 1/M^2$, so that, schematically,
\begin{equation}
    \delta f_{\mu\nu} \sim \frac{T_{\mu\nu}M^2}{1-\mu^2M^2}\,.
\end{equation}
Thus, the excitation of massive spin-2 QNMs is suppressed in the GR limit $\mu M\gg1$, as expected. These modes affect  the metric perturbations since, by replacing the above equation into Eq.~\eqref{eq:linear-Ricciflat-eom-g_matter}, in the $\mu M\gg1$ limit one obtains
\begin{equation}
    \delta G_{\mu\nu}= 8 \pi T_{\mu\nu}+{\cal O}(\mu^{-2} M^{-2})
\end{equation}
and hence
one recovers the ordinary GR result.

The above heuristic argument suggests that, in the GR limit, the extra modes will be present in the ordinary gravitational ringdown at ${\cal O}(\mu^{-2})={\cal O}(\alpha)$. 
However, within the EFT framework, through a field redefinition, it can be shown that that theory in vacuum is equivalent to GR to leading other in the quadratic-curvature couplings~\cite{Endlich:2017tqa}. Thus, an interesting question is whether some cancellation occurs and the new-mode excitation appears at ${\cal O}(\alpha^2)$, i.e. at the same order of the theories  with quartic terms in the curvature tensor, considered e.g. in\,\cite{Endlich:2017tqa, Cardoso:2018ptl}. This problem can be addressed by modelling through perturbation theory the actual excitation of the QNMs by a physical source, like e.g. an inspiralling particle.

In the interesting intermediate regime, $\mu M\sim 1$ (which is however not covered by the EFT), the amplitude of the massive spin-2 QNMs is not suppressed in $\delta g_{\mu\nu}$. In this case, the ringdown signal will contain extra modes that can be studied using the methods recently proposed in Ref.~\cite{Crescimbeni:2024sam}.

We postpone a detailed analysis of these interesting problems, as well as a comparison with the EFT prediction, to an ongoing follow-up work~\cite{follow-up}.

\section{Conclusions}
\label{sec:conclusions}
%
In this work we studied axial gravitational perturbations of BHs in quadratic gravity. This theory admits two distinct classes of static BH solutions: ordinary Schwarzschild BHs and hairy ones. The latter exist below a critical mass dictated by the coupling constant of the theory and have a significantly larger horizon radius (the solution space of static, rotating BHs has the same structure\,\cite{Sajadi:2023smm}).

We found numerical evidence that both Schwarzschild and hairy BHs are stable under non-radial, axial perturbations in the regime where radial instabilities are also absent.
Besides the usual deviations found in the QNMs of the hairy BH solution compared to the Schwarzschild one, we also found that the spectrum of Schwarzschild BHs in this theory is augmented by massive spin-2 modes, which are also present in the GW signal emitted in the ringdown stage of a binary BH coalescence. 
In the light of these findings, we have discussed the perspectives of finding observational constraints (or evidence) of quadratic gravity.

Future investigation will focus on the polar sector, which is expected to display the same qualitative features. Furthermore, it would be interesting to compute the signal emitted by inspiralling point particles, to quantify the amplitude of the extra massive spin-2 modes compared to the ordinary GR ones (see also\,\cite{OuldElHadj:2024psw} for some recent relevant results in massive gravity). 
This would also allow mapping the theory to the recently proposed theory-agnostic framework for extra non-GR modes in the ringdown~\cite{Crescimbeni:2024sam}.
Moreover, this would clarify the actual order of the effects of this theory in case it is considered in an EFT expansion.

Finally, here we focused on QNMs but the equation for massive spin-2 perturbations $\delta f_{\mu\nu}$ also admits quasi-bound state solutions with $\omega<\mu$~\cite{Brito:2013wya,Babichev:2013una,Brito:2020lup,Dias:2023ynv,East:2023nsk}. These modes have significant support only around the Bohr radius, $\sim 1/(M\mu^2)$~\cite{Brito:2015oca}, and exponentially decay at larger distances. Nevertheless, they might source ordinary gravitational perturbations and hence can contribute to the ringdown in a nontrivial way.

\acknowledgements
This work is partially supported by the MUR PRIN Grant 2020KR4KN2 ``String Theory as a bridge between Gauge Theories and Quantum Gravity'', by the FARE programme (GW-NEXT, CUP:~B84I20000100001), and by the INFN TEONGRAV initiative.
 We acknowledge financial support from the EU Horizon 2020 Research and Innovation Programme under the Marie Sklodowska-Curie Grant Agreement No. 101007855.

\appendix
\section{Background semi-analytical functions}\label{app:cfsolution}
Here we present the 4$^{th}$-order parametrization of the static, spherically symmetric hairy BH background derived in terms of continued fractions in~\cite{Kokkotas:2017zwt}. 
The spacetime metric is given by Eq.\,\eqref{eq:metric}, with
$A(r)$, $B(r)$ given in Eqs.\,\eqref{eq:fh} in terms of the functions $f(x)$, $h(x)$ of the variable $x=1-r_h/r$. These functions are:
\begin{align}
    f(x)=&1-\epsilon (1-x)-\epsilon(1-x)^2+\tilde{f}(x)(1-x)^3\,,\nonumber\\
    h(x)=&1+\tilde{h}(x)(1-x)^2\,,
\end{align}
where
\begin{align}
\tilde{f}(x)=\frac{\tilde{f}_1}{1+\frac{\tilde{f}_2 x}{ 1+\frac{\tilde{f}_3 x}{1+\frac{\tilde{f}_4 x}{1+\ldots}}}}\,,\\
\;\tilde{h}(x)=\frac{b_1}{1+\frac{\tilde{h}_2 x}{1+\frac{\tilde{h}_3x}{1+\frac{\tilde{h}_4 x}{1+\ldots}}}}\,,
\end{align}
\begin{align}
        \epsilon&=(1054 - 1203 p)\left(\frac{3}{1271} + \frac{p}{1529}\right)\,,\\
	\tilde{f}_1&=(1054 - 1203 p)\left(\frac{7}{1746}-\frac{5 p}{2421}\right)\,,\\
	\tilde{h}_1&=(1054 - 1203 p)\left(\frac{p}{1465}-\frac{2}{1585}\right)\,,\\
        \tilde{f}_2&=\frac{6 p^2}{17}+\frac{5 p}{6}-\frac{131}{102}\,,\\
        \tilde{h}_2&=\frac{81 p^2}{242}-\frac{109 p}{118}-\frac{16}{89}\,,\\
        \tilde{f}_3&=\frac{\dfrac{9921 p^2}{31}-385 p+\dfrac{4857}{29}}{237-223 p}\,,\\
	\tilde{h}_3&=-\frac{2 p}{57}+\frac{29}{56}\,,\\
        \tilde{f}_4&=\frac{\frac{9 p^2}{14}+\dfrac{3149 p}{42}-\dfrac{2803}{14}}{237-223 p}\,,\\
        \tilde{h}_4&=\frac{13 p}{95}-\frac{121}{98}\,,
\end{align}
$p=r_h\mu=r_h/\sqrt{2\alpha}$, and $\tilde{f}_{i}=\tilde{h}_{i}=0$ for $i>4$.
\section{Perturbations}
\label{ap:perturbations}
%
We consider linear perturbations of the metric field $g_{\mu\nu}$ and of the auxiliary field $f_{\mu\nu}$, around a static BH background:
\begin{align}
g_{\mu\nu}&=\bar{g}_{\mu\nu}+\varepsilon\, \delta g_{\mu\nu}\,,\\
f_{\mu\nu}&=\bar{f}_{\mu\nu}+\varepsilon\, \delta f_{\mu\nu}\,.
\end{align}
We consider perturbations with axial parity, expanding them in  tensorial spherical harmonics\,\cite{Regge:1957td,Maggiore:2018sht}:
\begin{align}
    (q^{\rm ax}_{\ell m})_{\mu\nu}=
    &
    \quad\sum_{\ell=1}^\infty \sum_{m=-\ell}^\ell\big[q_{\ell m}^{Bt}(\boldsymbol{t}_{\ell m}^{Bt})_{\mu\nu}+q_{\ell m}^{B1}(\boldsymbol{t}_{\ell m}^{B1})_{\mu\nu}\big]\nonumber\\
    &
    +\sum_{\ell=2}^\infty \sum_{m=-\ell}^\ell q_{\ell m}^{B2}(\boldsymbol{t}_{\ell m}^{B2})_{\mu\nu}\,,
\label{eq:axial_decomposition}
\end{align}
where
\begin{align}
   \boldsymbol{t}_{\ell m}^{Bt}=&
   \begin{pmatrix}
        0 & 0 & \csc\theta\,\partial_\varphi & -\sin\theta\,\partial_\theta\\
        0 & 0 & 0 & 0\\
        * & 0 & 0 & 0
   \end{pmatrix}Y_{\ell m}\,,\\[2mm]
   \boldsymbol{t}_{\ell m}^{B1}=&
   \begin{pmatrix}
        0 & 0 & 0 & 0\\
        0 & 0 & \csc\theta\,\partial_\varphi & -\sin\theta\,\partial_\theta\\
        0 & * & 0 & 0\\
        0 & * & 0 & 0
   \end{pmatrix}Y_{\ell m}\,,\label{eq:axial_basis_a}\\[2mm]\nonumber
\end{align}
are the axial vector harmonics, and
\begin{align}
   \boldsymbol{t}_{\ell m}^{B2}=&
   \begin{pmatrix}
        0 & 0 & 0 & 0\\
        0 & 0 & 0 & 0\\
        0 & 0 & -\csc\theta\,X & \sin\theta\,W\\
        0 & 0 & * & \sin\theta\, X
   \end{pmatrix}Y_{\ell m}\,,
 \label{eq:axial_basis_b}   
\end{align}
with $X=2\partial_\theta\partial_\varphi-2\cot\theta\,\partial_\varphi$, $W=\partial_\theta^2-\cot\theta\,\partial_\theta-\csc^2\theta\,\partial_\varphi^2$, are the axial tensor harmonics.

In order to fix the gauge of the perturbation expansion, we  consider an infinitesimal coordinate transformation $x'^\mu\to x^\mu+\xi^\mu$, which leaves the background invariant, and transform the metric perturbation as $h'_{\mu\nu}\to h_{\mu\nu}-2\nabla_{(\mu}\xi_{\nu)}$.
With an appropriate choice of
\begin{equation}
    \xi^{\rm ax}_\mu=\sum_{\ell=1}^\infty \sum_{m=-\ell}^\ell \Lambda_{\ell m}(t,r)\left(0,0,-\csc\theta\,\partial_\varphi,\sin\theta\,\partial_\theta\right)Y_{\ell m}
\end{equation}
we can set to zero the tensor massless perturbation (Regge-Wheeler gauge\,\cite{Regge:1957td}), i.e.
\begin{equation}
    \delta g^{\rm ax}_{\ell m}=
    \begin{pmatrix}
    0 & 0 & -h_0 \csc\theta\,\partial_\varphi & h_0 \sin\theta\,\partial_\theta \\
    0 & 0 & -h_1 \csc\theta\,\partial_\varphi & h_1 \sin\theta\,\partial_\theta  \\
    * & * & 0 & 0\\
    * & * & 0 & 0
    \end{pmatrix}
    Y_{\ell m} \, .
\label{eq:pert_ansatz_g}
\end{equation}
This choice does not affect the massive perturbations, which  have the general decomposition
\begin{equation}
\delta f^{\rm ax}_{\ell m}=
    \begin{pmatrix}
      0 & 0 & F_0\csc\theta\,\partial_{\phi} & -F_0\sin\theta\,\partial_{\theta}\\
      0 & 0 & F_1\csc\theta\,\partial_{\phi} & -F_1\sin\theta\,\partial_{\theta} \\
      * & * & -F_2\csc\theta\,X & F_2\sin\theta\, W  \\
      * & * & * & F_2\sin\theta\, X
    \end{pmatrix}
Y_{\ell m}\,.
\label{eq:pert_ansatz_f}
\end{equation}
Remarkably, in the massless limit $\mu\to0$ the decomposition\,\eqref{eq:pert_ansatz_f} does {\it not} reduce to the decomposition\,\eqref{eq:pert_ansatz_g}. Indeed, as discussed in\,\cite{Brito:2013wya}, in this limit the vector perturbations $F_0$, $F_1$ can be removed by a gauge choice.

\section{Numerical methods}
\label{ap:methods}
Here we briefly discuss the different numerical approaches employed to find the QNMs.
\subsection{Continued fraction method}
For the CF method we use the following ansatz~\cite{Pani:2013pma}
\begin{equation}
    \Psi^{(i)}\sim e^{-i \omega r_*} r^{\nu} e^{-q r}\sum_n a_n^{(i)} \psi_n(r)\,
\end{equation}
where $\Psi^{(i)}$ is any of the perturbation functions.

For $l\geq 2$ the axial perturbation functions corresponding to the massive mode, satisfy a pair of coupled differential equations, Eqs.\,\eqref{eq:Ricci_tensor_flat_system}. Substituting this ansantz leads to a three-term matrix-valued recurrence relation of the form
\begin{align}
\boldsymbol{\alpha}_0 \mathbf{U}_1+\boldsymbol{\beta}_0 \mathbf{U}_0&=0\,,\nonumber\\
\boldsymbol{\alpha}_n \mathbf{U}_{n+1}+\boldsymbol{\beta}_n \mathbf{U}_n+\boldsymbol{\gamma}_n \mathbf{U}_{n-1}&=0\,,\qquad n>0\,.
\end{align}
The quantity $\mathbf{U}_n=\left(a_n^{(1)},a_n^{(2)}\right)$ is a two-dimensional vectorial coefficient and $\boldsymbol{\alpha}_n$, $\boldsymbol{\beta}_n$, $\boldsymbol{\gamma}_n$ are $2\times 2$ matrices,
\begin{align}
&\boldsymbol{\alpha}_n =
\begin{pmatrix}
\alpha_n & 0 \\
   0     & \alpha_n 
\end{pmatrix}\,,\\[2mm]
&\boldsymbol{\beta}_n =
\begin{pmatrix}
\beta_n & \Lambda-2 \\
   -2   & \beta_n-3 
 \end{pmatrix}\,,\\[2mm]
&\boldsymbol{\gamma}_n =
\begin{pmatrix}
  \gamma_n & 6-3\Lambda \\
   0 & \gamma_n+9 
\end{pmatrix}\,,\\[2mm]
&\Lambda\equiv\ell(\ell+1)\, ,
\end{align}
with
\begin{align}
\alpha_n = & \; (n+1)(n+1-4i\omega)\,,\\[2mm]
\beta_n = & \; 2-\Lambda -2 \left(n^2+n-1\right)\nonumber\\
&+\frac{\omega ^2 (2 n-4 i \omega +1)}{q}\nonumber\\
&-3 q (2 n-4 i \omega +1)\nonumber\\
&+4 i (2 n+1) \omega -4 q^2+12 \omega ^2\,,\\[2mm]
\gamma_n = & \; q^{-2}[q^2 \left(n^2-4 i n \omega -6 \omega ^2-9\right)+q^4\nonumber\\
&+\omega ^4+2 q^3 (n-2 i \omega )-2 q \omega ^2 (n-2 i \omega )]\,. 
\end{align}
The matrix-valued three-term recurrence relation can be solved using matrix-valued continued fractions. The QNM frequencies are solutions to the equation $\mathbf{M}\mathbf{U}_0=0$, where
\begin{equation}
\mathbf{M}\equiv \boldsymbol{\beta}_0+\boldsymbol{\alpha}_0 \mathbf{R}^{\dagger}_0\,,
\end{equation}
with $\mathbf{U}_{n+1}=\mathbf{R}^{\dagger}_n \mathbf{U}_n$ and
\begin{equation}
\mathbf{R}^{\dagger}_n=-\left(\boldsymbol{\beta}_{n+1}+\boldsymbol{\alpha}_{n+1}\mathbf{R}^{\dagger}_{n+1}\right)^{-1} \boldsymbol{\gamma}_{n+1}\,.
\end{equation}
For nontrivial solutions we then solve numerically
\begin{equation}
\det |\mathbf{M}|=0\,.
\end{equation}

\subsection{Direct forward integration}
The forward integration technique may be used to search for the QNMs by incorporating a shooting method. Let us consider a system of $N$ coupled differential equations. The assumption of regularity of the  perturbation functions near the event horizon of the BH, leads to the asymptotic expansion:
\begin{equation}
\Psi^{(j)}=\;(r-r_h)^w\sum_{n=0} \psi_n^{(j)}(r-r_h)^{n+n_{(j)}}\,,\label{eq:bchor}
\end{equation}
where $j=1,\ldots N$ and the powers $n_{(j)}$ are specific to the particular perturbation function $\Psi^{(j)}$.
By imposing the field equation, we find that the boundary conditions at the horizon, Eq.\,\eqref{eq:bchor}, depend on $N$ independent variables. We then perform $N$ integrations from the horizon outwards by imposing the aforementioned boundary conditions. The perturbation function have the following asymptotic behaviour at infinity:
\begin{equation}
    \Psi^{(j)}\sim A^{(j)} e^{-i k r}r^{-\frac{q}{k}+m_{(j)}}+B^{(j)} e^{i k r}r^{\frac{q}{k}+m_{(j)}}\,,
\end{equation}
where $q$ depends on the equations at hand (e.g. on whether massive or massless modes are considered), and the powers $n_{(j)}$ are specific to the particular perturbation function $\Psi^{(j)}$.
Then, QNM solutions correspond to $B^{(j)}=0$.
We consider the following matrix
\begin{equation}
\boldsymbol{S} (\omega)=
 \begin{pmatrix}
  B_{1}^{(1)} & B_{1}^{(2)} & \ldots & B_{1}^{(N)} \\
  B_{2}^{(1)} & B_{2}^{(2)} & \ldots & \ldots \\
  \ldots  & \ldots  & \ldots & \ldots  \\
  B_{N}^{(1)} & \ldots & \ldots & B_{N}^{(N)}
 \end{pmatrix}\,,
\end{equation}
the components of which are the coefficients $B_i^{(j)}$. The superscript $j$ denotes a particular vector of the chosen basis, and corresponds to one of the different perturbation functions.
The subscript $i$ characterizes each one of the $N$ independent integrations.
The QNM frequency $\omega_0=\omega_R+i\omega_I$ will then correspond to the solutions of
\begin{equation}
\det|\boldsymbol{S}(\omega_0)|=0\,.
\end{equation}
\bibliography{bibnote}
\end{document}